\documentclass[apjl]{emulateapj}

\usepackage{natbib,graphicx}
\newcommand{\hone}{\hbox{\ion{H}{1}}}
\newcommand{\HI}{\hbox{\ion{H}{1}}}
\newcommand{\htwo}{\hbox{${\rm H_2}$}}
\newcommand{\coone}{\hbox{${\rm CO \ {\it J}=1-0}$}}
\newcommand{\cotwo}{\hbox{${\rm CO \ {\it J}=2-1}$}}

\newcommand{\hii}{\hbox{\ion{H}{2}}}

\newcommand{\Shone}{\hbox{$\Sigma_{\rm HI}$}}
\newcommand{\Shtwo}{\hbox{$\Sigma_{\rm H_2}$}}

\newcommand{\Sgas}{\hbox{$\Sigma_{\rm gas}$}}

\newcommand{\kms}{\hbox{$\rm km\,s^{-1}$}}
\newcommand{\Kkmsec}{\hbox{$\rm K \ km\,s^{-1}$}}

\newcommand{\kmsec}{\mbox{km\,s$^{-1}$}}

\newcommand{\msunpc}{\hbox{$\rm M_{\odot} \ pc^{-2}$}}

\submitted{Accepted by ApJ Letters}

\shorttitle{Atomic and Molecular Gas Surface Density}
\shortauthors{Wong et al.}

\begin{document}

\title{CARMA Survey Toward Infrared-bright Nearby Galaxies (STING). III. The Dependence of Atomic and Molecular Gas Surface Densities on Galaxy Properties}

\author{
Tony Wong\altaffilmark{1},
Rui Xue\altaffilmark{1},
Alberto D. Bolatto\altaffilmark{2},
Adam K. Leroy\altaffilmark{3},
Leo Blitz\altaffilmark{4},
Erik Rosolowsky\altaffilmark{5},
Frank Bigiel\altaffilmark{6},
David B. Fisher\altaffilmark{2},
J\"urgen Ott\altaffilmark{7},
Nurur Rahman\altaffilmark{8,9},
Stuart N. Vogel\altaffilmark{2},
\and
Fabian Walter\altaffilmark{10}
}

\altaffiltext{1}{Department of Astronomy, University of Illinois, 
Urbana, IL 61801, USA}
\altaffiltext{2}{Department of Astronomy, University of Maryland, 
College Park, MD 20742, USA}
\altaffiltext{3}{National Radio Astronomy Observatory, Charlottesville, 
VA 22903, USA}
\altaffiltext{4}{Department of Astronomy, University of California, 
Berkeley, CA 94720, USA}
\altaffiltext{5}{Department of Physics, University of Alberta, Edmonton, AB T6G 2E1, Canada}
\altaffiltext{6}{Institut f\"ur theoretische Astrophysik, Zentrum f\"ur Astronomie der Universit\"at Heidelberg, Albert-Ueberle Str. 2, 69120 Heidelberg, Germany}
\altaffiltext{7}{National Radio Astronomy Observatory, Socorro, NM 87801, 
USA}
\altaffiltext{8}{Department of Physics, C1 Lab 140, PO Box 524, University of Johannesburg, Auckland Park, Johannesburg 2006, South Africa}
\altaffiltext{9}{South Africa SKA Postdoctoral Fellow}
\altaffiltext{10}{Max-Planck-Institut f\"ur Astronomie, Konigstuhl 17, 
69117 Heidelberg, Germany}

\begin{abstract}

We investigate the correlation between CO and \HI\ emission in 18 nearby galaxies from the CARMA Survey Toward IR-Bright Nearby Galaxies (STING) at sub-kpc and kpc scales. 
Our sample, spanning a wide range in stellar mass and metallicity, reveals evidence for a metallicity dependence of the \HI\ column density measured in regions exhibiting CO emission.
Such a dependence is predicted by the equilibrium model of McKee \& Krumholz, which balances H$_2$ formation and dissociation.
The observed \HI\ column density is often smaller than predicted by the model, an effect we attribute to unresolved clumping, although values close to the model prediction are also seen.
We do not observe \HI\ column densities much larger than predicted, as might be expected were there a diffuse \HI\ component that did not contribute to H$_2$ shielding.
We also find that the H$_2$ column density inferred from CO correlates strongly with the stellar surface density, suggesting that the local supply of molecular gas is tightly regulated by the stellar disk.

\end{abstract}

\keywords{galaxies: ISM --- ISM: atoms --- ISM: molecules}

\section{Introduction}

Molecular gas in galaxies provides the raw material for star formation \citep[e.g.,][]{McKee:07}, yet in most galaxies the interstellar medium (ISM) is predominantly atomic. 
Understanding how the conversion of atomic to molecular gas is related to galaxy global properties, e.g., mass, morphology, and metallicity, is therefore critical to predicting the dense gas fraction and hence the star formation rate \citep[e.g.,][]{Lada:12}. 
Resolved studies of the spatial distribution of molecular and atomic gas over a wide variety of nearby galaxies can be used to test
proposed theories of ISM phase balance \citep[e.g.,][]{Krumholz:09a,Ostriker:10} and constrain the evolution of the ISM through cosmic time \citep[e.g.,][]{Obreschkow:09,Bauerm:10}. 

Over the past few decades, the 21\,cm \hone\ line and the 3\,mm \coone\ line have been the primary methods to study the atomic and molecular gas components of nearby galaxies.
The major reasons for using CO as a molecular gas tracer are its intrinsic brightness, modest excitation requirements, high abundance relative to \htwo, and accessibility to ground-based radio telescopes.
Thus, despite the uncertainty in converting CO intensity to \htwo\ column density, typically performed using a constant scaling factor (``X-factor''; see \citealt{Bolatto:13a} for a recent review), CO remains the preferred method to survey the molecular gas content in galaxies. 
In contrast to CO, the inference of atomic gas mass from 21\,cm emission is fairly direct, except in cases where the optical depth is thought to be significant. 
The primary challenge for systematically mapping \hone\ in nearby galaxies is achieving both high angular resolution and sensitivity.

While early observations \citep[e.g.,][]{Tacconi:86} had already indicated that \hone\ and CO are distributed very differently in galaxies, with CO largely confined to the stellar disk, the comparison of radial \hone\ and CO profiles across significant samples by \citet{Wong:02}, \citet[hereafter BR06]{Blitz:06}, and \citet{Leroy:08} revealed strong radial trends in the ratio of molecular to atomic hydrogen mass surface density,
\[R_{\rm mol} \equiv \Sigma_{\rm H_2}/\Sigma_{\rm HI}\,,\]
which exhibit a tight correlation with a simple estimate of the midplane gas pressure based on hydrostatic equilibrium, $P_{\rm mp} \propto \Sigma_{\rm gas}\sqrt{\rho_*}$.
This supported an earlier model by \citet{Elmegreen:93} that interstellar pressure (and, secondarily, radiation field) govern the \hone\ to \htwo\ transition.  
Note that $R_{\rm mol}$ is related to the quantity $f_{\rm mol} \equiv \Shtwo/\Sgas = R_{\rm mol}/(1+R_{\rm mol})$, which is more properly referred to as the molecular gas fraction but exhibits much less dynamic range. 

In recent years there has been a resurgence of interest in theoretical modeling of the \hone\ to \htwo\ transition, including attempts to model the physical and chemical processes on small scales in a time-dependent manner \citep{Pelupessy:06,Glover:10}, as well as construct analytic equilibrium models suitable for cloud scales and larger.  
The latter approach has been outlined in a series of papers by \citet{Krumholz:08}, \citet[hereafter KMT09]{Krumholz:09a}, and \citet[hereafter MK10]{McKee:10}.  
Roughly speaking, an equilibrium \htwo\ abundance arises from a competition between \htwo\ formation on grains and photodissociation by UV radiation.  
Thus, for a given cloud the molecular gas fraction depends on the dust optical depth of the cloud (since dust excludes dissociating radiation and encourages molecule formation) and the strength of the external radiation field.  
A dependence of $f_{\rm mol}$ on metallicity enters via the correlation of metallicity with the dust abundance.  
Already there have been attempts to apply the empirical $R_{\rm mol}$--$P_{\rm mp}$ relation of BR06 and the theoretical predictions for $f_{\rm mol}$ of KMT09 and MK10 to cosmological contexts \citep[e.g.,][]{Obreschkow:09,Fu:10,Lagos:11,Kuhlen:12}, underscoring the importance of testing their applicability across a broader range of galaxies.

The CARMA\footnote{The Combined Array for Research in Millimeter-wave Astronomy (CARMA) is operated by the Universities of California (Berkeley), Chicago, Illinois, and Maryland, and the California Institute of Technology, under a cooperative agreement with the University Radio Observatory program of the National Science Foundation.} Survey Toward IR-Bright Nearby Galaxies (STING) is a major effort to measure the molecular gas properties and their relation to star formation in a sample of $\sim$20 nearby star-forming galaxies. 
Unlike the precursor BIMA Survey of Nearby Galaxies (SONG; \citealt{Helfer:03}), the STING sample is specifically chosen to cover a wide range in stellar mass, and thus covers a wide range in associated galaxy properties (star formation rate, color, luminosity, metallicity, etc.).
Supplementing STING CO data with ancillary multi-wavelength data from the ultraviolet (UV) to infrared (IR), \citet{Rahman:11} studied the spatially resolved star formation law and its sensitivity to different analysis methods in NGC~4254, and have recently extended this study to the full STING sample \citep{Rahman:12}. 
The molecular gas data from the STING survey also permit a study of atomic-to-molecular gas transition, which is the subject of this Letter. 
We compare the STING \coone\ dataset with archival \hone\ data from the NRAO Very Large Array (VLA) to examine the atomic and molecular gas relation in 18 galaxies of the STING sample on sub-kpc to kpc scales.
Our principal observational result is a clear dependence of \hone\ column density on metallicity, and an equally striking dependence of \htwo\ column density on stellar surface density.
A more detailed presentation of these data and comparison with theoretical models will be presented in a future work (Xue et al.\ 2014, in preparation, hereafter X14).

\begin{deluxetable*}{lccccccccccccccccc}
\tabletypesize{\tiny}
\tablecaption{Sample Properties\label{tbl:galprops}}
\tablewidth{0pt}
\tablecolumns{11}
\tablehead{
\colhead{Name} & 
\colhead{Dist.} &
\colhead{$i$} &
\colhead{P.A.} &
\colhead{$M_B$} &
\colhead{12+log(O/H)} &
\colhead{Metal Ref.} &
\colhead{$b_{\rm maj}\times b_{\rm min}$ (P.A.)} &
\colhead{$r_{\rm phy}$} &  
\colhead{$\sigma$\Shone} &
\colhead{$\sigma$\Shtwo} \\
\colhead{} & 
\colhead{(Mpc)} &
\colhead{($^{\circ}$)} & 
\colhead{($^{\circ}$)} & 
\colhead{(mag)} &
\colhead{(dex)} &
\colhead{} &
\colhead{($\arcsec\times\arcsec$)~($^{\circ}$)} &
\colhead{(kpc)} &
\colhead{(\msunpc)} &
\colhead{(\msunpc)}
}
\startdata
NGC \phn337 & 21.3 & 52 & 121 & $-19.99\pm0.24$ & $8.60\pm0.34$ & 1 & $28.5\times20.5$\, (195) & \phn4.74 & 0.09 & 0.61 \\
NGC \phn772 & 30.2 & 37 & 315 & $-21.57\pm0.49$ & $8.87\pm0.13$ & 6 & $48.1\times43.0$\, (\phn88) & \phn8.50 & 0.04 & 0.67 \\
NGC 1156 & \phn7.0 & 52 & 272 & $-17.73\pm0.34$ & $8.18\pm0.10$ & 3 & $\phn7.0\times\phn6.2$\, (336) & \phn0.38 & 0.40 & 2.97 \\
NGC 1569 & \phn2.5 & 63 & 112 & $-17.65\pm0.96$ & $8.13\pm0.12$ & 3 & $\phn6.6\times\phn5.6$\, (\phn17) & \phn0.18 & 0.25 & 1.01 \\
NGC 1637 & \phn9.8 & 39 & 213 & $-18.63\pm0.56$ & $8.80\pm0.34$ & 4 & $23.3\times19.2$\, (169) & \phn1.34 & 0.08 & 0.94 \\
NGC 2782 & 16.1 & 30 & \phn75 & $-18.79\pm0.97$ & $8.59\pm0.10$ & 3 & $\phn7.5\times\phn6.2$\, (\phn87) & \phn0.61 & 0.57 & 4.48 \\
NGC 2976 & \phn3.6 & 65 & 335 & $-17.20\pm0.50$ & $8.67\pm0.31$ & 1 & $\phn8.9\times\phn7.7$\, (\phn46) & \phn0.37 & 0.12 & 0.98 \\
NGC 3147 & 40.9 & 32 & 147 & $-21.72\pm0.37$ & $9.02\pm0.36$ & 2 & $21.3\times18.0$\, (136) & \phn4.40 & 0.10 & 1.12 \\
NGC 3198 & 14.0 & 72 & 215 & $-19.90\pm0.28$ & $8.62\pm0.28$ & 1 & $\phn7.7\times\phn6.5$\, (\phn66) & \phn1.53 & 0.10 & 0.77 \\
NGC 3486 & 15.6 & 36 & 263 & $-19.99\pm0.59$ & $8.75\pm0.32$ & 2 & $59.2\times49.8$\, (\phn79) & \phn4.69 & 0.07 & 0.37 \\
NGC 3593 & \phn5.5 & 67 & \phn90 & $-16.92\pm0.08$ & $8.29\pm0.26$ & 2 & $15.4\times14.6$\, (346) & \phn1.06 & 0.03 & 0.37 \\
NGC 4151 & \phn6.6 & 21 & \phn22 & $-17.71\pm1.40$ & $8.41\pm0.28$ & 2 & $\phn9.6\times\phn8.8$\, (\phn66) & \phn0.33 & 0.59 & 3.22 \\
NGC 4254 & 15.6 & 31 & \phn69 & $-20.67\pm0.23$ & $8.79\pm0.34$ & 1 & $10.6\times\phn9.6$\, (\phn95) & \phn0.89 & 0.40 & 1.54 \\
NGC 4536 & 14.7 & 68 & 301 & $-19.74\pm0.37$ & $8.61\pm0.40$ & 1 & $10.1\times\phn9.4$\, (\phn81) & \phn1.86 & 0.14 & 0.72 \\
NGC 4605 & \phn5.5 & 67 & 300 & $-17.85\pm0.41$ & $8.43\pm0.28$ & 2 & $\phn5.5\times\phn4.9$\, (\phn72) & \phn0.36 & 0.30 & 1.24 \\
NGC 4654 & 16.1 & 62 & 125 & $-20.02\pm0.28$ & $8.83\pm0.27$ & 5 & $16.9\times16.4$\, (\phn55) & \phn2.82 & 0.06 & 0.45 \\
NGC 5371 & 27.7 & 48 & \phn11 & $-20.93\pm0.55$ & $8.90\pm0.34$ & 2 & $85.1\times45.6$\, (\phn81) & 16.68 & 0.04 & 0.11 \\
NGC 5713 & 21.4 & 33 & 203 & $-19.95\pm0.53$ & $8.64\pm0.40$ & 1 & $16.9\times16.2$\, (328) & \phn2.08 & 0.14 & 0.77 \\
NGC 6951 & 23.3 & 46 & 138 & $-21.55\pm0.27$ & $8.99\pm0.36$ & 2 & $11.1\times\phn6.5$\, (\phn48) & \phn1.83 & 0.37 & 2.53
\enddata
\tablecomments{Metallicity references: 1: \citet{Moustakas:10}; 2: \citet{Moustakas:10} ($L$-$Z$); 3: \citet{Engelbracht:08}; 4: \citet{vanZee:98}; 5: \citet{Pilyugin:04b}, \citet{Dors:06}; 6: \citet{Anderson:10}}
\end{deluxetable*}

\section{Observations and Data Reduction}

The STING sample is composed of northern ($\delta > -20\arcdeg$), moderately inclined ($i < 75\arcdeg$) galaxies within 45 Mpc selected from the IRAS Revised Bright Galaxy Survey \citep{Sanders:03} to sample a wide range of stellar masses (and consequently, ISM metallicity).
CO observations were conducted in the C and D configurations of CARMA to yield a robustly weighted synthesized beam of 3\arcsec--4\arcsec\ across a field of view extending to roughly half the optical radius.
Calibration was performed using the MIRIAD package, but imaging and deconvolution were performed using CASA\footnote{Common Astronomy Software Applications, http://casa.nrao.edu} \citep{McMullin:07}, taking advantage of the multi-scale {\sc clean} feature of CASA to better recover extended structures by searching for model components on multiple scales (in our case, point sources as well as Gaussians twice and five times the width of the synthesized beam).
For four galaxies (NGC 3198, 4254, 4536, and 5713) with \cotwo\ single-dish maps from the HERACLES project (\citealt{Leroy:09}, \citealt{Leroy:13}), a detailed comparison in X14 suggests that the CARMA maps recover $\gtrsim$75\% of the flux within the field of view.

We searched the NRAO Data Archive\footnote{\url{http://archive.nrao.edu}} for public \hone\ 21\,cm data of STING galaxies obtained in B, C, or D configurations of the VLA with adequate velocity resolution ($\lesssim$20 \kms) to separate line from continuum. 
Archival data were found for 19 out of the 23 galaxies, so the present study is restricted to this sub-sample. 
In addition, NGC 3593 shows unusually low deprojected \HI\ column densities, which we attribute to its high inclination and significant optical depth in the 21\,cm line, so we do not include it in the present analysis.
To ensure the most uniform possible data processing, we reprocessed the visibility data using CASA.
We employed multi-scale {\sc clean} for well-resolved galaxies (with map resolution $<1$ kpc), and flagged regions with strong continuum emission ($T_b>50$ K) as potentially susceptible to optical depth effects.
Further details on the pipeline as well as data sources can be found in X14.
The FWHM beam parameters of the resulting \hone\ maps (major and minor axis and position angle) are given in Table~\ref{tbl:galprops}, and set the limiting resolution for our analysis.

\begin{figure*}
\center
\includegraphics[scale=.75]{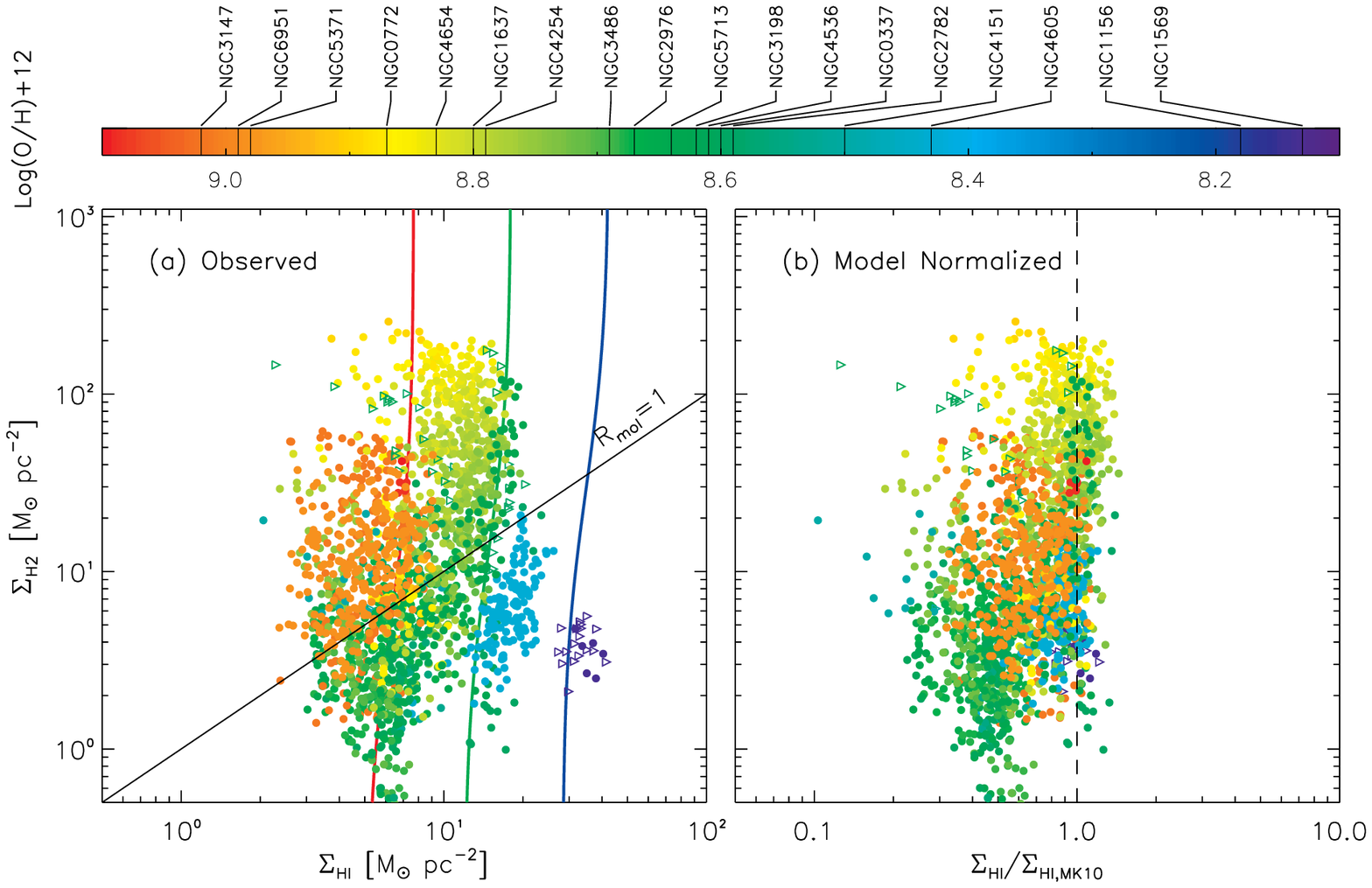}
\includegraphics[scale=.75,trim=0 0 0 80, clip=true]{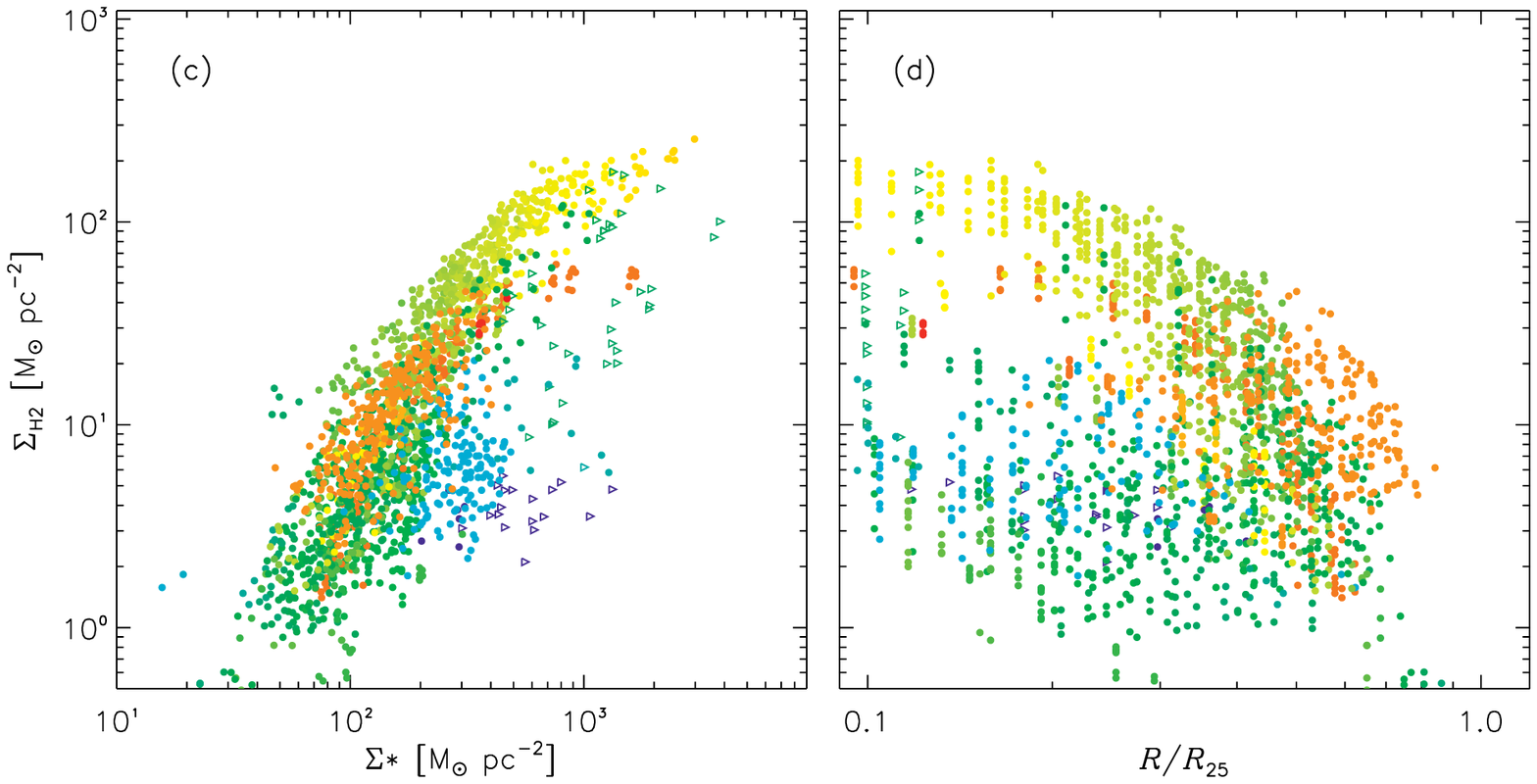}
\caption{(a) Comparison of \hone\ and \htwo\ surface densities sampled at the resolution of the \hone\ data.  Points are color coded by gas-phase metallicity, assumed uniform except for four galaxies (NGC 1637, 3198, 4254, and 4654) where a measured gradient has been applied.  A diagonal line corresponds to the locus of equal \hone\ and \htwo\ surface densities.  The correlation of characteristic \hone\ column density with metallicity is apparent as a color gradient in the plotted points.  The MK10 predictions for three different metallicities ($12+\log({\rm O/H})=9.1$, 8.65, and 8.2) are shown as colored lines.  (b) The same data and color coding but with the abscissa normalized by the predicted value from MK10 as appropriate to each point.  Panels (c) and (d) compare \Shtwo\ with stellar surface density and normalized galactocentric radius respectively.
\label{fig:h1h2_hires}}
\end{figure*}

A smoothing and masking process was used to generate integrated line intensity images from the CO and \hone\ data cubes.
The cube was first smoothed along the spatial and velocity axes to increase the signal-to-noise ratio.
A mask was then generated from the smoothed cube by starting at emission peaks defined by at least two consecutive $>$4$\sigma$ channels (where $\sigma$ is the RMS noise in the smoothed cube) and expanding outward to a 2$\sigma$ edge.
This mask was then applied to the original cube, blanking regions outside the mask, before summing the velocity channels.
Finally, the integrated intensity images were deprojected to face-on using the orientation parameters given in Table~\ref{tbl:galprops} after being regridded to a common pixel scale and smoothed to a common resolution (the latter chosen to achieve a circular beam when deprojected).
For reference, Table~\ref{tbl:galprops} also gives, as a measure of the physical resolution, the length scale $r_{\rm phy}$ corresponding to the deprojected \hone\ beam using the median redshift-independent distance (e.g., from the Tully-Fisher method) reported by NED.\footnote{The NASA/IPAC Extragalactic Database (NED) is operated by the Jet Propulsion Laboratory, California Institute of Technology, under contract with the National Aeronautics and Space Administration.}

Face-on mass surface densities were calculated by using the optically thin assumption for \hone\ and a CO-to-H$_2$ conversion factor of $\alpha_{\rm CO} = 4.4$ \msunpc\,(\Kkmsec)$^{-1}$ for CO \citep[e.g.,][]{Leroy:08}.
Reported surface densities include a factor of 1.36 to account for helium.
The typical 1$\sigma$ uncertainty in the deprojected surface density, based on a typical integration window of 20 \kmsec\ for CO and 40 \kmsec\ for \hone, is given in Table~\ref{tbl:galprops} and is appropriate for maps smoothed to the \hone\ resolution.
The deprojected intensity maps were sampled onto a hexagonal grid with a spacing between grid centers equal to the (deprojected) FWHM resolution of the image, following the methodology of \citet{Leroy:13}.

\section{Results}

Figure~\ref{fig:h1h2_hires}(a) displays a point-by-point comparison of \Shone\ and \Shtwo\ at the limiting resolution of the \hone\ data.  
Only regions where both \hone\ and CO are detected at the $>$3$\sigma$ level are shown.  
Open triangle symbols indicate \hone\ measurements which are potentially contaminated by absorption in front of bright continuum emission, and a diagonal line labeled $R_{\rm mol}$=1 divides \hone\ and \htwo-dominated regimes.
Points are color coded by metallicity, determined as follows.
For six objects lacking spectroscopic gas-phase metallicities (NGC 3147, 3486, 4151, 4605, 5371, and 6951), we derive a fiducial metallicity for the galaxy by applying the luminosity-metallicity relation of \citet{Moustakas:10}.
For four galaxies (NGC 1637, 3198, 4254, and 4654) for which a metallicity gradient has been determined from multiple spectroscopic measurements, we determine a local metallicity appropriate to each grid sample.
For the remaining galaxies, we assign a single metallicity value (given in Table~\ref{tbl:galprops}) across the galaxy, based on published \hii\ region spectroscopy.
(Since the gradients, when measured, are typically around $-0.5$ dex per $R_{25}$, and our observations are confined to $R \ll 0.5\,R_{25}$, inclusion or omission of a gradient has little effect on our results.)
We find that points for a given galaxy tend to be distributed vertically, consistent with a nearly constant value of \Shone\ when CO is detectable.
However, horizontal offsets between galaxies are clearly apparent, with lower metallicity galaxies shifted to the right.

In Figure~\ref{fig:h1h2_hires}(b), \Shone\ has been normalized by the \hone\ surface density predicted by MK10 (their Equations 13, 86, 91, and 93) for the observed \Shtwo\ and metallicity; this is rather insensitive to the actual value of \Shtwo\ (and thus the adopted CO-to-\htwo\ conversion factor), as indicated by the nearly vertical shape of the model curves drawn in Figure~\ref{fig:h1h2_hires}(a).
For the sample as a whole we find a reduced dispersion in \Shone\ (from 0.23 to 0.17 dex) after applying the MK10 prediction, as well as a shift of some of the lower metallicity outliers into the main concentration of points.
We therefore see evidence for the metallicity dependence predicted by the model.
Nevertheless, the distribution is strongly skewed to the left of the dashed line, where observed and predicted column densities are equal.
Thus, the model frequently overpredicts \Shone\ by factors of up to 5, especially for lower values of \Shtwo.  
The trends and scatter we observe remain unchanged when we repeat the analysis for 9 galaxies at a fixed deprojected resolution of 2 kpc, suggesting they are not strongly influenced by varying $r_{\rm phy}$ across our sample.

\begin{figure*}
\center
\includegraphics[scale=.75]{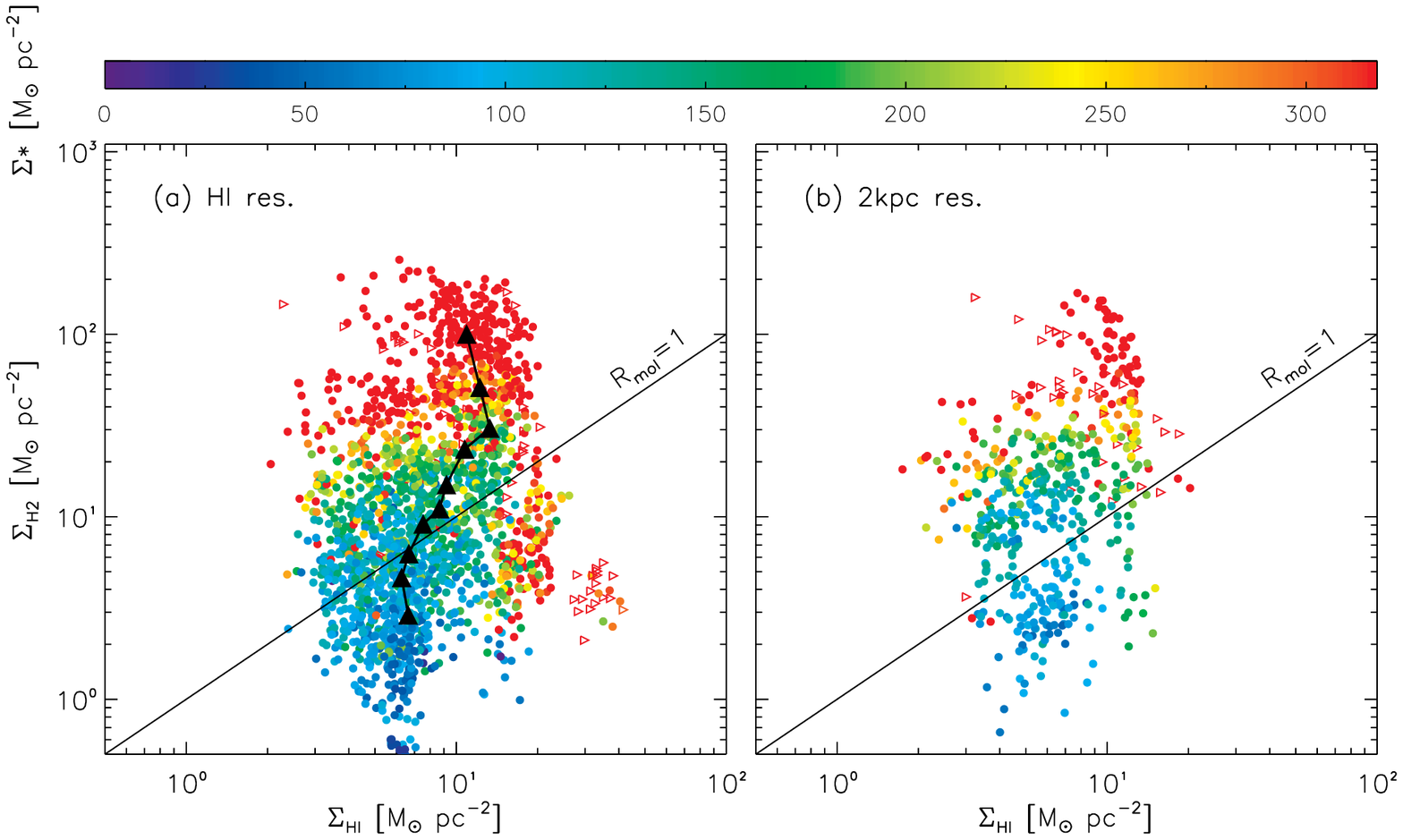}
\caption{(a) The same data points are shown as in Figure~\ref{fig:h1h2_hires}(a), but the points are color coded by stellar surface density as inferred from {\it Spitzer} 3.6\,$\mu$m surface brightness.  
Solid triangles denote average values of \Shone\ and \Shtwo, calculated in deciles of $\Sigma_*$.
There is a strong gradient in stellar surface density along the \Shtwo\ axis.
(b) \Shtwo\ vs. \Shone, again color coded by $\Sigma_*$, determined at a fixed resolution of 2 kpc for galaxies with $r_{\rm phy}<2$ kpc.
\label{fig:h1h2_stars}}
\end{figure*}

The tendency for the MK10 model to overpredict \Shone\ may be linked to a reduction in measured peak gas column densities (relative to what would be measured on GMC-like scales) as a result of spatial averaging over kpc-scale beams.
Because $f_{\rm mol}$ is a steeply increasing function of \Sgas\ in the model, the suppression of peaks in \Sgas\ by beam averaging causes the model to underestimate the molecular fraction and thus overestimate \Shone.
Higher values of \Shtwo\ may be subject to less beam dilution and may better reflect the conditions under which the MK10 model is directly applicable (i.e., a resolved GMC complex and its atomic envelope).
This may account for the tendency for \Shone/$\Sigma_{\rm HI,MK10}$ to approach unity for larger \Shtwo.
Further discussion of this effect can be found in KMT09.

To investigate trends in \Shtwo\ independent of \Shone, Figures~\ref{fig:h1h2_hires}(c) and (d) plot values of \Shtwo\ against stellar mass surface density ($\Sigma_*$) and scaled galactocentric radius, with points still color-coded by metallicity.  
The stellar mass surface density is inferred from {\it Spitzer} 3.6\,$\mu$m imaging (S$^4$G project; \citealt{Sheth:10}) using $\Sigma_{*}=317.7\times I_{3.6}$ \citep{Eskew:12}, where $I_{3.6}$ is the background-subtracted 3.6\,$\mu$m intensity in units of MJy sr$^{-1}$.
A strong correlation between \Shtwo\ and $\Sigma_*$ is apparent, whereas no relationship with $R/R_{25}$ is discernible in the sample as a whole.

Figure~\ref{fig:h1h2_stars} provides another view of the \Shtwo--$\Sigma_*$ correlation by showing the same data as Figure~\ref{fig:h1h2_hires}(a), but with points color coded by $\Sigma_*$.
At the full \hone\ resolution (left panel), it is clear that \Shtwo\ correlates well with $\Sigma_*$, except for a few low-metallicity, \hone-dominated galaxies where \Shtwo\ is low despite high $\Sigma_*$.
(Such exceptions may reveal inaccuracies in our assumption a uniform X-factor and uniform mass-to-light ratio in the IR.)
Average values of \Shone\ and \Shtwo\ (filled triangles), calculated in uniformly spaced logarithmic bins of $\Sigma_*$, underscore the much greater sensitivity of \Shtwo\ to $\Sigma_*$.
The same trend is seen at a fixed resolution of 2 kpc (right panel), for the 9 galaxies observed at sufficiently high resolution, except that several of the low-metallicity systems drop out of the sample.
We stress that the CO measurements are taken over areas much larger than an individual GMC, so the observed trend probably originates from changes in the covering fraction of GMCs rather than the actual surface density of GMCs.

\section{Discussion and Summary}

We have compared the CO and \hone\ surface brightnesses in individual apertures across 18 galactic disks from the CARMA STING sample spanning a range of metallicity and stellar mass.  For simplicity we have plotted only independent quantities, avoiding intrinsic correlations between axes and assumptions about which combination of observables is most physically relevant for star formation or ISM physics.
We also exclude non-detections (upper limits) from our analysis, so we are sensitive only to trends in CO-detected regions, which may be quite distinct from trends related to the detectability of CO \citep[e.g.,][]{Saintonge:11a}.

Our principal result is a clear dependence of the the characteristic \hone\ column density on metallicity, as predicted by the models of KMT09 and MK10.
A similar result had been obtained for a dwarf galaxy sample by \citet{Fumagalli:10}, but with less of a clear metallicity dependence.
We find that a major success of the MK10 model is its ability to predict the upper envelope of the observed \Shone\ values, although the full distribution of points appears skewed below the predicted values by unresolved clumping within the beam.
In particular, we do not see evidence for a diffuse \hone\ component that does not contribute to \htwo\ shielding and would therefore cause the observed \Shone\ to exceed the model prediction.
On the other hand, the fact that the predicted value of \Shone\ is sometimes reached in low \Shtwo\ regions suggests that \hone\ and CO need not suffer equally from beam dilution.
This would imply a difference in how \hone\ and CO are spatially distributed, as found by recent studies of the SMC \citep{Bolatto:11} and other galaxies \citep{Leroy:13a}, and be inconsistent with the bulk of the \HI\ being confined to the GMC envelopes modeled by MK10.

The correlation we observe between \Shtwo\ and $\Sigma_*$ (setting aside low-metallicity systems for which the CO-to-\htwo\ conversion factor or 3.6\,$\mu$m mass-to-light ratio may change) is hardly unexpected: a correspondence between CO and stellar radial profiles has been noted in many previous studies \citep[e.g.,][]{Regan:01,Leroy:08,Leroy:09,Schruba:11,Bigiel:12}.
Still, a comparison of Figures 1(a) and 2(a) vividly illustrates that the \hone\ and \htwo\ column densities of galactic disks appear to be governed by distinct processes.
It is possible that the two trends we observe, lower \Shone\ with higher metallicity (which often associates with higher $\Sigma_*$) and higher \Shtwo\ with higher $\Sigma_*$, reinforce each other to produce a tight correlation of $R_{\rm mol}$ with $\Sigma_*$ and estimates of hydrostatic pressure.
However, since $\Sigma_*$ is dominated by the old stellar population, the processes that maintain its tight correlation with \Shtwo\ remain unclear.

Studies probing GMC size scales find that the mass surface densities of GMCs are confined to a fairly narrow range \citep{Bolatto:08,Roman:10b}, implying that kiloparsec-scale variations in \Shtwo\ are largely reflecting the CO areal covering factor.  
We therefore hypothesize that GMCs are more abundant in regions of galaxies with higher $\Sigma_*$.  
A simple explanation is that star formation in GMCs is responsible for building up the stellar disk, resulting in a stellar disk that naturally mimics the \htwo\ distribution.
However, observational evidence for GMC lifetimes of only a few tens of Myr \citep{Kawamura:09,Murray:11} raises the question of why the GMCs should be regenerated at the same locations in the disk.  
In present-day, highly evolved galaxies with gas fractions $\sim$0.1, it seems more plausible to argue that the \htwo\ distribution is governed by the stellar distribution.
For instance, in the thermal equilibrium model of \citet{Ostriker:10}, higher stellar density leads to higher pressure in the diffuse gas (as required by hydrostatic equilibrium), increasing the cooling rate and leading to the formation of bound clouds.
Alternatively, \htwo\ consumption could be regulated by $\Sigma_*$-dependent gravitational instabilities \citep[an idea explored further by][]{Zheng:13}, or a significant fraction of the total \htwo\ supply may be recycled from the old stellar disk \citep[an idea explored further by][]{Leitner:11}.

\acknowledgments

We thank the referee for suggestions which have improved the presentation of results.
T. W. and R. X. acknowledge support from NSF University Radio Facilities Program and the NASA Astrophysics Data Analysis Program.
Support for CARMA construction was derived from the Gordon and Betty Moore Foundation, the Eileen and Kenneth Norris Foundation, the Caltech Associates, the states of California, Illinois, and Maryland, and the NSF. 
Funding for ongoing CARMA development and operations are supported by NSF and CARMA partner universities.  
The National Radio Astronomy Observatory is a facility of the National Science Foundation operated under cooperative agreement by Associated Universities, Inc.

{\it Facilities:} \facility{VLA}, \facility{CARMA}, \facility{$Spitzer$}

\end{document}